\documentclass[preprint,runinaddress,frontmatterverbose,aps, showpacs,prb,floatfix]{revtex4-1}
\usepackage{graphicx}
\usepackage{dcolumn}
\usepackage{bm}
\usepackage{graphicx}
\usepackage{color}
\usepackage{colortbl}
\usepackage[dvipsnames]{xcolor} 
\usepackage{tabularx}
\usepackage{epsfig}
\usepackage{amsmath}
\usepackage{amssymb}
\usepackage{graphicx}
\usepackage{dcolumn,epstopdf}
\usepackage{bm}
\usepackage{wasysym}
\usepackage{times}
\definecolor{grn}{rgb}{0,0,0.54}

\usepackage[colorlinks,bookmarks=false,citecolor=blue,linkcolor=red,urlcolor=blue]{hyperref}
\usepackage{latexsym}
\usepackage{amssymb}
\usepackage{amsfonts}
\usepackage{amsmath}
\def\be{\begin{equation}}
\def\ee{\end{equation}}
\newcommand{\bea}{\begin{eqnarray}}
\newcommand{\eea}{\end{eqnarray}}

\begin{document}

\title{NMR Investigation of the Low Temperature Dynamics of solid $^4$He doped with $^3$He impurities }

\author{S. S. Kim$^1$\footnote{Current address: \'Ecole Polytechnique F\'ed\'erale de Lausanne, Lausanne, Switzerland}}
\author{C. Huan$^1$\footnote{Current Address: Georgia Institute of Technology, Atlanta, Ga, USA}}
\author{L. Yin$^1$}
\author{J. S. Xia$^1$}
\author{D. Candela$^2$}
\author{N. S. Sullivan$^1$}
\affiliation{$^1$Department of Physics \& National High Magnetic Field Laboratory, University of Florida, Florida 32611, USA.\\
$^2$Department of Physics, University of Massachusetts, Amherst, Massachusetts 01003, USA}
\email{sullivan@phys.ufl.edu}

\begin{abstract}
	The lattice dynamics of solid $^4$He has been explored using pulsed NMR methods to study the motion of $^3$He impurities in the temperature range where experiments have revealed anomalies attributed to superflow or unexpected viscoelastic 
properties of the solid $^4$He lattice.  We report  the results of measurements  of the nuclear spin-lattice and spin-spin relaxation times that measure the fluctuation spectrum at high and low frequencies, respectively, of the $^3$He motion that results from quantum tunneling in the $^4$He matrix. The measurements were made for $^3$He concentrations $16<x_3<2000$ ppm.
	For $^3$He concentrations  $x_3$ = 16 ppm and 24 ppm, large changes are observed for both the spin-lattice relaxation time $T_1$ and the spin-spin relaxation time $T_2$ at temperatures close to those for which the anomalies  are observed in measurements of torsional oscillator responses and the shear modulus. These  changes in the NMR relaxation rates were not observed for higher $^3$He concentrations.
	
\end{abstract}
\pacs{67.80.bd, 67.30.hm, 76.60.-k}
\keywords{Suggested keywords}
\date{\today}
\maketitle
\footnote{\dag  THE}
\section{Introduction}	The observation of distinctive anomalies in the frequency and dissipation of torsional oscillators containing solid $^4$He, often referred to as  non-classical rotational inertia fractions (NCRIFs), by Kim and Chan
 \cite{ISI:000188068100037, ISI:000224136000041} 
has stimulated enormous activity. This is  because the NCRIFs could be the signature of a supersolid state as outlined by Leggett\cite{PhysRevLett.25.1543}  and predicted by Andreev and Lifshitz several years ago.\cite{AndreevandLifshitz}
A large number of independent experiments \cite{PhysRevLett.100.065301, PhysRevLett.97.165301, Reppy2010, Davis2011, Kojima08}  have shown that the NCRIF magnitude and temperature dependence are strongly dependent on defects such as $^3$He impurities\cite{Gumann2011,Kojima12} and the quality of the crystals, and can be made very small by very careful annealing. \cite{ISI:000224136000041}  The observation of NCRIFs was preceded by reports of a striking anomaly in the measurement of sound attenuation in solid $^4$He by Goodkind and colleagues who had compared their observations to that predicted for a true phase transition of a fraction of the solid to a supersolid state.\cite{Goodkind1,Goodkind2}  Additional support for the interpretation of the NCRIF anomalies in terms of a true thermodynamic phase change to a new state of matter was provided by the observation of a small but distinct contribution to the heat capacity of the solid at the same temperatures as those  reported for the onset of the NCRIFs.\cite{Linetal2007}

A straightforward interpretation in terms of a transition of part of the solid to a coherent superfluid component has, however, been hampered by a number of puzzling observations. These include the lack of evidence for a critical exponent, an apparent very low critical velocity\cite{PhysRevLett.97.165301}, the absence of experimental evidence for fourth sound modes \cite{Aokietal,Kwonetal} 
 and the null results of attempts to observe pressure- induced superflow though small restrictions,\cite{DayBeamishPRL06}  although  mass transport has been observed with the use of porous vycor glass conduits used to allow mass flow to the solid via superfluid in the pores.  \cite{HallockPRL08,HallockPRB09}
	In addition,  measurements of the shear modulus of solid $^4$He by Beamish and colleagues \cite{PhysRevLett.104.195301,PhysRevB.79.214524} have revealed a prominent frequency dependent change in the elastic shear modulus with an enhanced dissipation peak having a temperature dependence comparable to that observed for the NCRIF.  These results suggest that the torsional oscillator anomalies may result from unusual elastic\cite{ElasticPropsBeamish09} or viscoelastic\cite{Dorsey09} properties of the solid $^4$He rather than superfluidity.  Other interpretations attribute the anomalies to macroscopic superflow mediated by defects or dislocation networks\cite{Boninsegni06} or in terms of a vortex model\cite{P.W.Anderson} for which the high temperature tail of the NCRIF is associated with the finite response time of vortices to the oscillating flow fields in the TOs. Another set of researchers suggest that the superfluidity arises from non-equilibrium behavior leading to superflow along defects or the formation of a quantum ``superglass" around extended defects\cite{Balatsky09} with ultra-slow relaxation dynamics reminiscent of glass dynamics.\cite{Graf2011} 

The need for any interpretation to account for both sample dependent coherent non-inertial   flow and changes in the lattice elastic properties was confirmed strikingly by  Kim {\it et al.} \cite{DKimetal2011} who observed that the TO response could be changed significantly by rotating the cryostat, with the resonant frequency changing with speed, but that at the same time, the TO mode that showed significant drive dependence was not susceptible to changes in the elastic modulus of the lattice. \cite{DKimetal2011}  

It is clear that  the  dynamics of the $^4$He lattice plays an important role in the low temperature bulk properties of solid $^4$He and rather than observing a phase transition to a supersolid state one may be observing a complex thermally excited dynamical response, or both, in overlapping temperature ranges.
	It is therefore important to study the microscopic dynamics of  solid $^4$He  using different techniques and especially non-invasive techniques since all previous studies have involved the application of macroscopic external mechanical stimuli. Measurements of the quantum tunneling of the  $^3$He atoms in the solid provide a unique means of probing microscopic lattice properties because the  $^3$He -$^4$He exchange rates  depend exponentially on the lattice separation and on the magnitude and dynamics of the  crystal field deformation that surround impurities.\cite{Eshelby1956, Landesman1975137} The characteristic NMR relaxation rates are determined by the modulation of the nuclear dipole-dipole interactions by the tunneling motion   and the scattering of the diffusing atoms by the crystal deformation field around the $^3$He impurities and other lattice defects.\cite{Landesman1975137, PhysRevB.11.3374}  The NMR relaxation rates are therefore very sensitive to the local elastic properties of  solid $^4$He and to any changes in the crystal ground state that would modify the tunneling rate. We have therefore carried out systematic measurements of the nuclear spin relaxation rates of $^3$He impurities in solid $^4$He for a wide range of concentrations with an emphasis on low concentrations $x_3<30$ ppm for which one expects well characterized NCRIFs from previous studies and yet sufficient to obtain good signal to noise ratios for NMR measurements. Also for these concentrations, the amount of the $^3$He  localized on dislocations estimated from the expected density of dislocations for well-annealed samples\cite{Iwasa2010}  is expected to be a small fraction of the total, less than 1 ppm, so that one can be sure that the experiments probe the  motion of the $^3$He atoms through bulk solid $^4$He. Preliminary results of these NMR studies have been reported elsewhere.\cite{SSKimJLTP2010,SSKimPRL2011}

\section {Experimental Arrangement}

The NMR cell was designed as a nested cross-coil arrangement with receiving and transmitting coils orthogonal to each other and the applied magnetic field (see Fig.\,1 ). 
The inner coil is a cylindrical receiving coil wound around a polycarbonate sample cell which contained an {\it in situ} Straty-Adams pressure gauge.  The outer coil is a thermally isolated orthogonal transmitting coil for the RF pulses that slides onto the receiving coil. This nested arrangement\cite{Huan}   provides (i) minimization of the unwanted pick-up of the transmitter pulse by the receiving coil, and (ii) adequate thermal isolation  from the heat generated in the transmitting coil that is heat sunk to the still  of the dilution refrigerator. 
Thermalization of the sample was assured by thermal contact with a silver post extending from the dilution refrigerator and the temperature was measured using a Cernox resistance thermometer calibrated against a $^3$He melting curve thermometer.  The sample gas was admitted via a capillary sealed with epoxy (stycast 2585 GT) at the opposite end of the cell. The sample cell could withstand pressures up to 100 bar and remain superfluid leak-tight. 

\begin{figure}[!hb]
\includegraphics*[width=.52\textwidth]{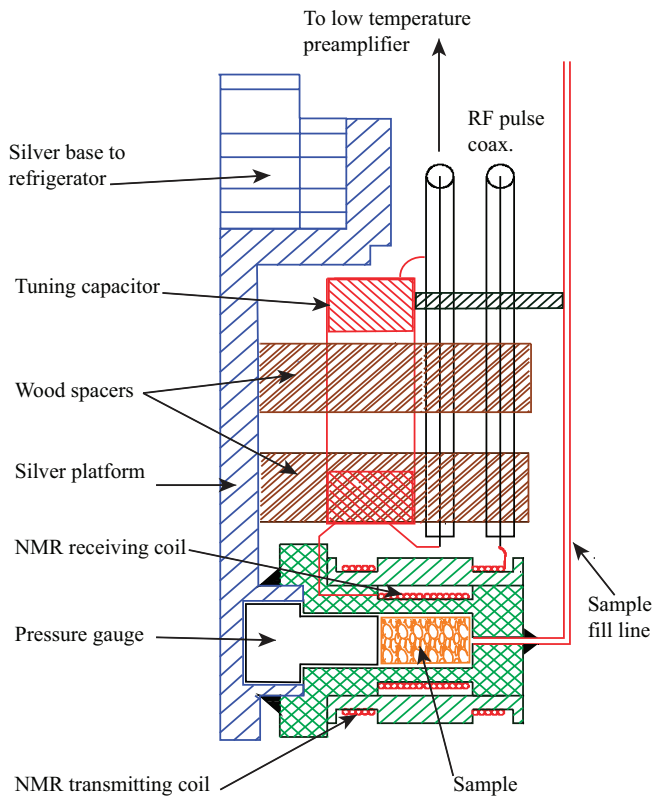}
\caption{ \label{fig: NMR cell design}
	(Color online) Schematic representation of the low temperature NMR cell. The sample in this NMR cell is cylindrical, 5.0 mm diameter and 7.5 mm long.
	The preamplifier and tuning capacitor are located on a 1~K cold plate located a distance of 1.2~m from the sample cell.
	The RF transmitting and receiving coils are simplified in this figure.
}
\end{figure}



Samples were grown by the blocked capillary method. Gas samples of pre-determined $^3$He fraction were prepared at room temperature by mixing pure (99.99\%) $^3$He and $^4$He, compressing the gas mixture to about 50 bar and filling the sample cell with the gas mixture to a pressure of 46 bar, after which the cell was cooled to 1.2 K. When the temperature is cooled to the melting curve, the helium pressure follows the melting curve while the mixture solidifies, after which the pressure measured by the gauge in the cell remains constant indicating that the sample is solid. The samples were prepared to all have the same final pressure of $27.75\pm 0.05$ bar at 1.2 K. The molar volume of the  samples were determined to be $V_m=20.85 \pm 0.05 $ cm$^3$ from the PVT data of Grilly and Mills\cite{GrillyMills} and the formula derived by Mullin.\cite{Mullin} Finally the samples were annealed for 24 hours just below the melting point except for one case for which a minimum annealing of  0.5 hours was carried out to determine the effect of the crystal quality on the nuclear spin relaxation times.

The measurements of the  nuclear spin-spin and spin-lattice relaxation times were carried out using a superheterodyne pulsed NMR spectrometer operating at  a Larmor frequency of 2.05MHz. At this frequency the  calculated  relaxation times due to exchange motion were of the order of $10^4$ s. at $^3$He concentrations $x_3=20$ ppm, extrapolating from previous measurements.\cite{GRZ,Grigoriev1973,Schratter1984,Greenberg1972} 
The longitudinal relaxation time is expected to increase rapidly with frequency\cite{GRZ}, making studies at higher Larmor frequencies extremely difficult.

The signal/noise ratio for samples with $x_3 \approx 20$ ppm is very weak $(<10^{-2})$  for a 10 kHz bandwidth employing a standard geometry with resonant NMR circuit connected directly to  a room temperature amplifier. In order to enhance the signal/noise we developed a  preamplifier that could be operated at low temperatures in the applied magnetic field.\cite{Huan} The device used a pseudomorphic high electron mobility field effect transistor with adjustable bias so that the output could be matched to a 50 ohm cable. The total power dissipation was 0.5 mW and to attain the lowest sample temperatures the amplifier was separated from the sample cell by a 75 cm length of cable and anchored to a 1K thermal plate. A typical NMR echo recorded using a  $90_x-180_x$ RF pulse sequence for a 500 ppm $^3$He sample is shown in Fig.\,2 after signal averaging for 10 pulse sequences using a 17 kHz bandwidth. As it is necessary to wait several times $T_1$ between pulse sequences, up to 15 hours were required to obtain each data point at the lowest temperatures.  Therefore the total number of data points is modest, but sufficient we believe to reveal and confirm relevant features in the temperature dependence of the relaxation times.

\begin{figure}[h]
\includegraphics*[width=0.52\textwidth]{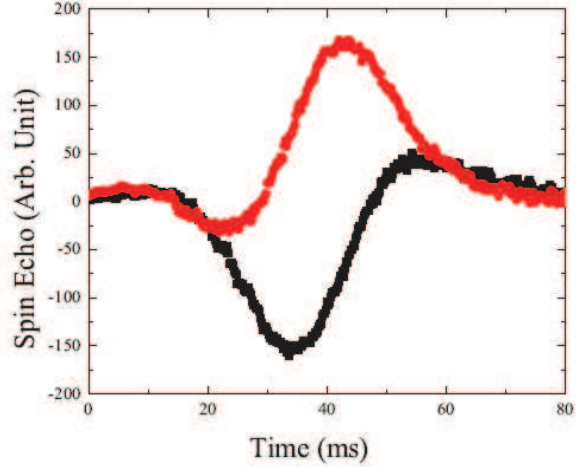}
\caption{ \label{fig: NMR Echo}
	(Color online) Example of NMR Hahn echo for a sample of 500 ppm $^3$He in solid $^4$He at T=0.4K, showing the in-phase and out-of-phase components.}

\end{figure}

One of the most important properties of the sample that must be determined is the exact $^3$He concentration, because for the long capillary line used (1.2 m below the 1K plate) an appreciable fraction of the $^3$He atoms can plate out on the capillary walls and lead to changes in the $^3$He concentration of the solid sample compared to that of the original gas concentration. Fortunately, the NMR signal itself allows one to obtain an accurate calibration  by measuring the NMR echo amplitude for fixed spectrometer gain at intermediate temperatures and comparing to a relatively high concentration sample (2000 ppm) for which any fractional change would be small. In Fig.\,3 we show the temperature dependence of NMR echo amplitudes for several different samples. All the samples show the characteristic Curie law behavior until reaching temperatures of the order of 0.1K where phase separation occurs.  At these low temperatures the $^3$He forms Fermi liquid droplets with temperature independent NMR amplitudes and with different relaxation times. Using this procedure one can  measure the $^3$He concentration to approximately 5\% accuracy.
For example, for a sample prepared from a nominally 30 ppm gas mixture, the final solid $^3$He fraction was measured to be 24 ppm.
Note that a  small deviation was observed in the amplitude of the echo signal for the 16 ppm sample at $T=175$ mK and this is attributed to the sudden and unexpected sharp increase in $T_1$ at that temperature. Waiting times between measurements were set at 5 times $T_1$ to provide ~5\% accuracy in amplitude measurements but at 175mk this drops to 2.5 times $T_1$ and a small decrease in observed amplitude results.

\begin{figure}[h]
\includegraphics*[width=0.55\textwidth]{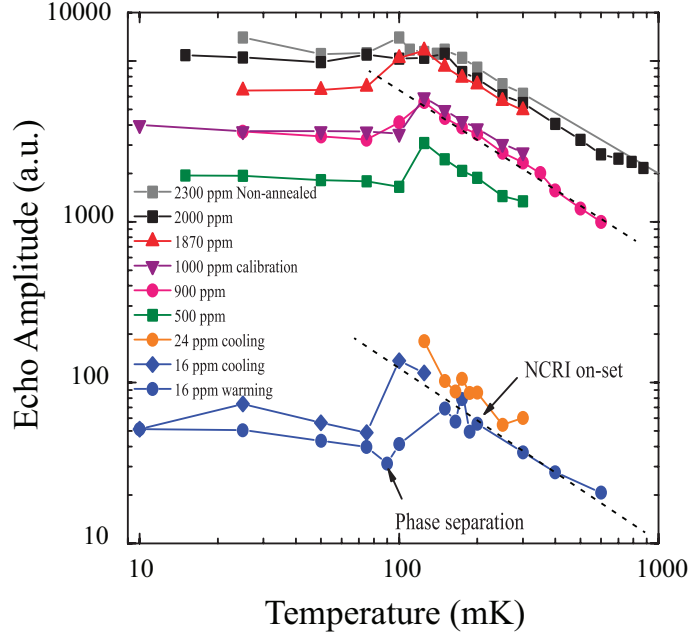}
\caption{ \label{fig: Ampli Several samples}%
	(Color online) Observed temperature dependence of NMR echo amplitudes for several different samples measured for the same spectrometer pulse and gain settings. The broken lines represents the Curie-law behavior.  }
\end{figure}

\section{Experimental Results}

\begin{figure}[h]
\includegraphics*[width=0.61\textwidth]{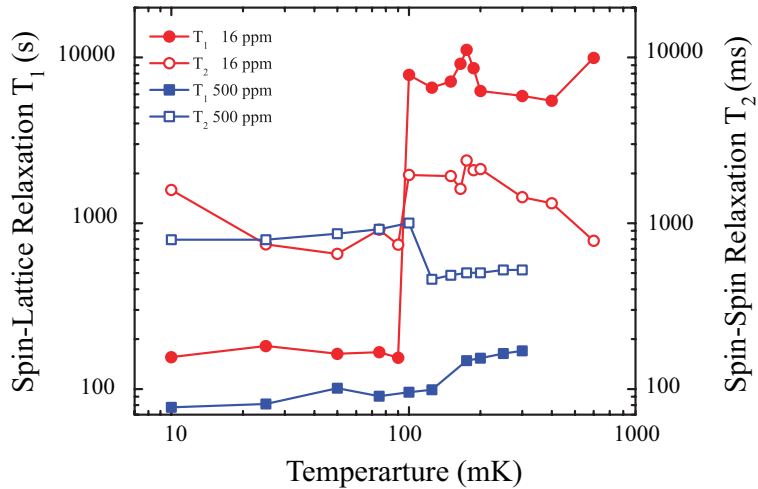}
\caption{ \label{fig: T1andT2 2 samples}
	(Color online) Observed temperature dependence of the spin-lattice and spin-spin relaxation time for two samples (i)  $x_3$=16 ppm, and (ii) $x_3$=500 ppm.  }
\end{figure}

Fig.\,4  shows the observed temperature dependence of the nuclear spin-lattice relaxation time ($T_1$) and the spin-spin relaxation time ($T_2$) for a  sample with $x_3$= 16 ppm and the comparison with the results for a sample with $x_3=$ 500 ppm. 
The temperature independent tunneling induced relaxation occurs for $0.2<T<0.55$ K and a small upturn  in $T_1$  is observed for $T> 0.55$ K with a corresponding downturn in $T_2$. This high temperature behavior is attributed to the onset of the effect of thermally activated vacancies.  The most interesting features are the sharp peak in $T_1$ at $T=175$ mK  and a less well-defined minimum in $T_2$ at the same temperature for the sample with $x_3$=16 ppm. The same behavior as a function of temperature is observed for a sample with $x_3$ =24 ppm prepared with a considerably reduced annealing time. \cite{SSKimPRL2011}.
The sharp changes in $T_1$ and $T_2$ are not observed for samples with concentrations $x_3 >$ 200 ppm. These features near 175 mK are unexpected in the traditional interpretations of the relaxation of dilute $^3$He in solid $^4$He that attribute the relaxation to a temperature independent quantum tunneling $J_{34}$ of the $^3$He atoms through the $^4$He lattice. The temperatures at which these  sharp features for $T_1$ and $T_2$ occur coincide with the temperatures for which anomalies are observed for measurements of the rotational inertia\cite{ ISI:000224136000041,ISI:000188068100037}  and shear modulus experiments\cite{PhysRevB.79.214524,Day_Beamish}. On cooling below 175 mK, a  precipitous drop in the value of $T_1$ is observed  at $T= 95$ mK which is  the temperature associated with the phase separation of the solid mixture into $^3$He rich droplets  in otherwise pure solid $^4$He. This interpretation is confirmed by the observed change in the NMR amplitude below the phase separation temperature as shown in Fig. 4.  and by the significant hysteresis on cycling through the phase separation temperature. \cite{SSKimPRL2011}


\section{Discussion}
The theoretical treatments of the relaxation of dilute $^3$He atoms in solid $^4$He\cite{Landesman1975137,ZhEhsp.T,LandesmanWinter,PhysRevB.11.3374,LandesmanVacancy}  all take into consideration the lattice deformation surrounding a  $^3$He impurity due to the increased zero-point motion of the $^3$He atoms with respect to the lattice compared to the $^4$He atoms\cite{NSandAL} The resulting deformation has been described in terms of a long range anisotropic interaction between $^3$He atoms given by
\begin{equation}
 K(r_{ij})=K_0(3cos^2\theta_{ij}-1) (\frac{a_0}{r_{ij}})^{3}
\end{equation}

\noindent for the deformation energy at site j due to an impurity at site i, where $a_0$ is the lattice constant.\cite{Eshelby1956,Slyusarev1977}. The $^3$He atoms travel through the lattice and scatter from one another with a closest approach given by the distance $b_c$ for which the kinetic energy of the tunneling particle ($\sim J_{34}$) becomes comparable to the elastic deformation energy 
\begin{equation} b_c= (\frac{J_{34}}{K_0})^{1/3}a_0.\end{equation}
 The mean separation of the $^3$He atoms $r_m=a_0x_3^{-1/3}$ and for $x_3 \geq x_{30}=\frac{J_{34}}{K_0} \sim 10^{-3}$ the $^3$He atoms are in continuous interaction with elastic fields of one another. For $x_3 < x_{30}$, the atoms move coherently by tunneling until scattered by other $^3$He atoms or other lattice defects.
This simple estimation of the concentration $x_{30}$ for which there is a cross-over from simple coherent diffusion to the continuous interaction regime was analyzed in detail by Landesman and Winter\cite{LandesmanWinter}   using a moment expansion for the calculation of the physical diffusion and they found a very different value  with $x_{30} \sim (\frac{J_{34}}{K_0})^2 \sim 10^{-6}$. Huang {\it et al.}\cite{PhysRevB.11.3374} re-examined this estimate using a more precise accounting of of the energetics of the scattering process and estimated $x_{30} \sim 10^{-4}$. As we will show below, the Landesman model gives  a good description of the values of the spin lattice relaxation time measured by different research groups  with a characteristic concentration dependence of $T_1 \propto x_3^{-2/3}$ for $x_{3} > 10^{-4}$ and a very different concentration dependence below 10 ppm.

In order to understand the possible origins of the experimentally observed peak in {\it T$_1$} at low temperatures we need to examine how the tunneling motion of the $^3$He atoms and their accompanying lattice deformations determine the nuclear spin-lattice relaxation. Landesman\cite{Landesman1975137}  treated the motion of the $^3$He atoms in the presence of the deformations using a fictious spin model. In this model a fictious spin $S_i$ takes the values $0$ or $\pm 1$ according to whether a site i is occupied by a $^4$He atom or a $^3$He atom with real nuclear spin $I_j^z =\pm1$, respectively. The probability that a  site j is occupied by a $^3$He atom is represented by $\tau_j=(S_Z^j)^2$ and the elastic deformation energy by the Hamiltonian 
\begin{equation}H_K= -2\hbar\sum_{jk}K_{jk}\tau_j\tau_k.
\end{equation}
The tunneling Hamiltonian is given by
\begin{equation}H_T=-2\hbar J_{34}\sum_{j,k} P_{jk}
\end{equation}
 where $P_{jk}= \frac{1}{2}(S_j^zS_j^{+}S_k^{-}S_k^z + S_j^zS_j^{-}S_k^{+}S_k^{z}) + h.c$  is the permutation operator for atoms at sites i and j.

The time dependence of the fictious spin operators $S_j$ and $\tau_j$ are given by 
\begin{equation}S_j^{+}(t)=e^{iH_Kt}S_j^{+}e^{-iH_Kt}=S_j^{+}e^{i\omega_jt}
\end{equation} where $\omega_j=2\sum_{j<k}K_{jk}\tau_k$. In this time dependence Landesman did not include a lattice relaxation term to account for relaxation of the lattice as the deformation surrounding the impurity atom moves through the lattice. Given the recent results from elastic studies  of solid helium at low temperatures this could be an important effect. The relaxation times can now be estimated by calculating the correlation functions of the nuclear dipole-dipole interactions. We consider the reduced correlation function \begin{equation}\Gamma_{ijkl}^m = \frac{4}{x_3} \langle T_{ij}^m(t)T_{kl}^m(0)^ \dag \rangle 
\end{equation} where the $T_{ij}$ are the irreducible nuclear spin operators (that transform analogously to the spherical harmonics $Y_2^m(\Omega_{ij}$)). $T_{ij}^m(t)=e^{iH_Mt}T_{ij}^me^{-iH_Mt}$ with $H_m=H_K+H_T$ the total Hamiltonian responsible for the motions in the lattice. For the dilute lattice the tunneling Hamiltonian does not commute with all $T_{ij}$ and one finds that the derivative 
\begin{equation}\ddot \Gamma_{ij,ij}(t)/\Gamma(0) = - 12x_3J_{34}^2\sum_pF_{ip}(t)cos[(K_{ij}-K_{jp})t] (\frac{a_0}{r_{ij}})^6
\end{equation} 
with $F_{ip}(t) =\prod_s cos(\omega_s(t)$ where $\omega_s=x_3(K_{is}-K_{ps})t$. Evaluating $F(t)$ using the statistical method of Anderson{\cite{Anderson1951} yields 
\begin{equation}\ddot \Gamma(t)/\Gamma(0) =-48\Lambda^{'}x_3J_{34}^2 exp[-x_3\Lambda(K_0t)^{3/4}]\end{equation} where $\Lambda{'}=0.28$ and $\Lambda=8.77$.\cite{LandesmanWinter,Landesman1975137} 
 Integrating Eq'n (8) Landesman\cite{Landesman1975137} finds a correlation time $\tau_c$ given by 
\begin{equation} \tau_c^{-1}= B\frac {J_{34}^2}{K_0} x_3^{-1/3}\end{equation} with $B=23$.

\begin{figure}[h]
\includegraphics*[width=0.55\textwidth]{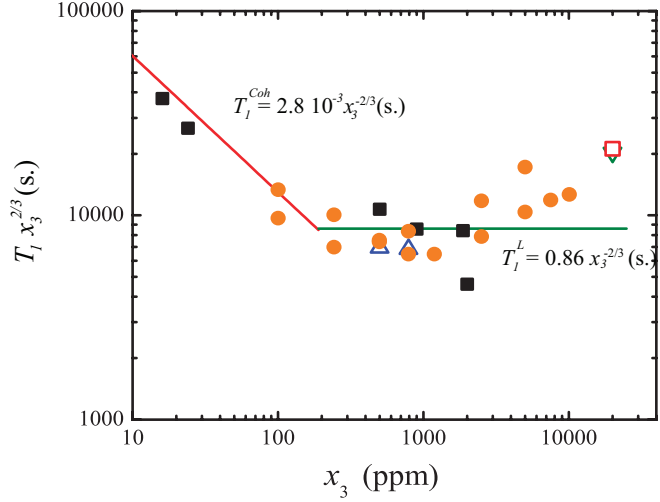}
\caption{ \label{fig: Amplitude includ 16 ppm}
	(Color online) Observed $^3$He concentration dependence of the nuclear spin-lattice relaxation time for dilute $^3$He in solid $^4$He. Orange circles, Schratter {\it et al.} \cite{Schratter1984}, up triangles, Allen {\it et al.} \cite{1982JLTP...47..289A}, diamonds Schuster {\it et al.} \cite{Schuster1996}, down triangles \cite{Hirayoshi}, open squares \cite{Greenberg}, solid squares \cite{SSKimPRL2011}. }
\end{figure}

The correlation time $\tau_c <\omega_L^{-1}$ for all the experiments reported and we therefore find for the calculated nuclear spin-lattice relaxation time in Landesman's model \begin{equation}T_1^L =\frac{\omega_L^2}{46M_2}\frac{1}{J_{eff}}x_3^{-2/3}\end{equation} where $M_2$ is the second moment for pure $^3$He and $J_{eff} = J_{34}^2/K_0$.
As shown in Fig.\,5 this calculated value provides a good description of the observed values of $T_1$ in the temperature-independent "plateau" region for $^3$He concentrations $x_3 >100$ ppm for $\frac{J_{eff}}{2\pi}=1.2$ kHz and no other adjustable parameter. At lower concentrations a much stronger concentration dependence is observed with $T_1\propto x_3^{-4/3}$. This behavior at low concentrations is attributed to the crossover from the continuous interaction regime to a region of coherent diffusion for which the characteristic time  is determined by the time for a $^3$He atom to travel the mean distance $r_m=x_3^{-{1/3}}a_0$  between $^3$He atoms. This time is given by $\tau_{coh}=r_m/v_g$ where the group velocity $v_g=a_0zJ_{34}$. We find  $\tau_{coh}=(zJ_{34})^{-1}x_3^{1/3}$ leading to a calculated longitudinal relaxation time $T_1^{coh}=2.8 \times 10^{-3}x_3^{4/3}$ shown by the solid red line of Fig.\,5 for $J_{34}/(2\pi)$ = 1.2 MHz. 


The predicted temperature independent relaxation for $x_3$ = 16 ppm is  shown by the broken  lines in Fig.\,6. While the fit to experimental data is good at high temperatures the anomalous  relaxation observed at 175 mK is more than a factor of two larger than the Landesman prediction for a simple tunneling motion of the $^3$He atoms through the lattice. The anomaly occurs well above the phase separation temperature (90 mK) and is not attributable to the phase separation which would lead to a sharp decrease in the relaxation time and not the sharp peak that was observed. This conclusion was confirmed by studying a  sample with 24 ppm that was never never cooled to the phase separation temperature (see Fig.\,6).  

It is important to note that a simple change in the elastic field interaction $K$  associated with thermal population of excited states with different values for $K_0$ would not explain the observed results and would simply lead to a broad step in the value of $T_1$. The results therefore imply that the anomalous peak observed for $T_1$ results either from a change in the dynamical properties of the lattice that occur at 175 mK or that there is a phase transition that occurs at that temperature and the peak observed in $T_1$  results from the fluctuations at the phase transition.

\begin{figure}[h]
\includegraphics*[width=0.55\textwidth]{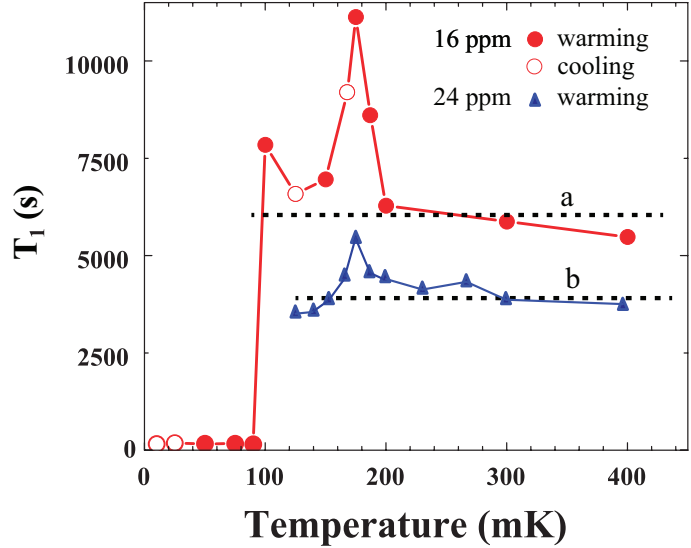}
\caption{ \label{fig: T1 data function of x}
	(Color online) Comparison of the values of the observed nuclear spin-lattice relaxation times  for $x_3$=16 ppm (red circles)  and $x_3$=24 ppm (blue triangles) with the theoretical model of Landesman  (broken lines a and b) for tunneling $^3$He impurities in the $^4$He lattice with a fixed lattice distortion. The sharp drop at 90 mK marks the well known phase separation with the formation of pure $^3$He nanodroplets. The 24 ppm sample was never cooled below 120 mK to avoid hysteresis and memory effects associated with the phase separation.
	}
\end{figure}

\begin{figure}[h]
\includegraphics*[width=0.57\textwidth]{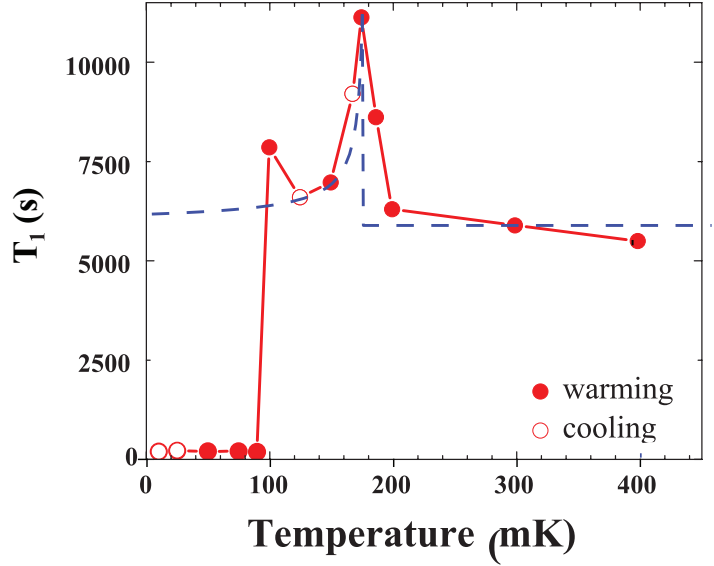}
\caption{ \label{fig: Super fit to T1 function of T}
	(Color online) Comparison of the observed temperature dependence of the nuclear spin-lattice relaxation time $T_1$ with the dependence expected for a true phase transition at T=175 mK for a sample with $x_3$ = 16 ppm as described in the text.
	 }
\end{figure}

\subsection{Effect of a transition to supersolid state on the NMR relaxation times}

We now consider what would be expected if there was a simple phase transition to a state in which there was a small fraction of superfluid condensate, and then we will discuss how the effect of a dynamical relaxation of the lattice as described by Beamish and colleagues would influence the nuclear spin lattice relaxation time $T_1$.  

If there was a sharp phase transition at a fixed temperature $T_C$ below which  a superfluid state appears one would expect a sudden change in the spectral density at $T_C$ that determines $T_1$ and $T_2$ because of the sharp increase in fluctuations just below $T_C$ that would modulate the spin-spin interactions (assuming that the $^3$He atoms are carried by the superfluid component). One of the most significant results of Chan {\it et al.}\cite{ISI:000224136000041}  is that the critical velocity attributed to the superfluid interpretation of the NCRIFs is very low, typically $10^{-4}$ m/s. The fluctuations would therefore add weight to the spectral density at low frequencies and not at high frequencies ($>$MHz). Because the spectral densities are determined by the fluctuations of the $^3$He dipole-dipole interactions (and no other magnetic interactions are present) they are  normalized and thus the increase in spectral weight at low frequencies would result in a decrease in the spectral weight at high frequencies with an increase in $T_1$ and a concommittant decrease in $T_2$.

If we designate the component of the spectral density due to $^3$He particle exchange as $J_E(\omega)$ and the component due to a possible superfluid component as $J_S(\omega)$, we have 
\begin{equation}\int\frac{1}{T_1}d \omega = \int [J_E(\omega)+J_S(\omega)]d\omega=\pi M_2 \end{equation} where $M_2$ is the NMR second moment. For liquid $^4$He the coherence length scales as $\xi=a_0(T_C-T)^{-2/3}$ where $T_C$ is the critical temperature, and because of the very low critical velocities (inferred from torsional oscillator measurements), we anticipate a characteristic relaxation time that behaves as $\tau_S=\tau_{so}(T_C-T)^{-2/3}$ for $T<T_C$, using $\tau_{SO}\cong r_m/v_c$, where $r_m=x^{-1/3}a_0$ is the mean separation between atoms, $a_0$ is the lattice spacing and $v_c$ is the critical velocity.
If the supersolid component component has weight $\alpha(T)$ then the exchange component must have weight $1-\alpha(T)$  so that the sum rule $\int \frac{1}{T_1}=\pi M_2 $ is obeyed. Thus the observed relaxation rate will be given by
\begin{equation} \frac{1}{T_{1Obs}}=\alpha(T)J_S(\omega_L)+(1-\alpha(T))J_E(\omega_L)\end{equation} where $J_S(\omega_L)$ is negligible. If $T_{1E}$ is the calculated value of $T_1$ for exchange motion only, we find
\begin{equation}\frac{T_{1Obs}}{T_{1E}}= \frac{1}{[1-\alpha(T)]}\end{equation} where for a supersolid fraction we take $\alpha(T)=a_1\frac{T_0}{|T-T_C)|^{2/3}}$ for $T<T_C$ and $\alpha(T)=0$ for $T>T_C$. $a_1$ is an adjustable parameter related to the "`supersolid"' density which is of the order of 0.01 but sample dependent.

An example of a fit using the above expression for $T_{1Obs}$ is shown by the broken line in Fig.\,7 for $T_C$= 175 mK and $a_1=$ 0.01 with no other adjustable parameter. In reality we would expect to have a small Gaussian spread  (of about 7 mK) in the critical temperature because of the inhomogeneities and it would not be difficult to obtain a good fit to $T_1$ for reasonable numbers but no other experiment points to a sharp transition (except for the onset of NCRIFs in very pure samples), e.g.  no sharp peak in the heat capacity is reported. A more stringent test of this interpretation is provided by examining the results for the $T_2$ measurements for the same sample. The same argument as used for evaluating $T_1$ needs to be followed for $T_2$ but this time we need to know the values for $\tau_{s0}$ and the exchange time $\tau_E$.

\begin{figure}[h]
\includegraphics*[width=0.57\textwidth]{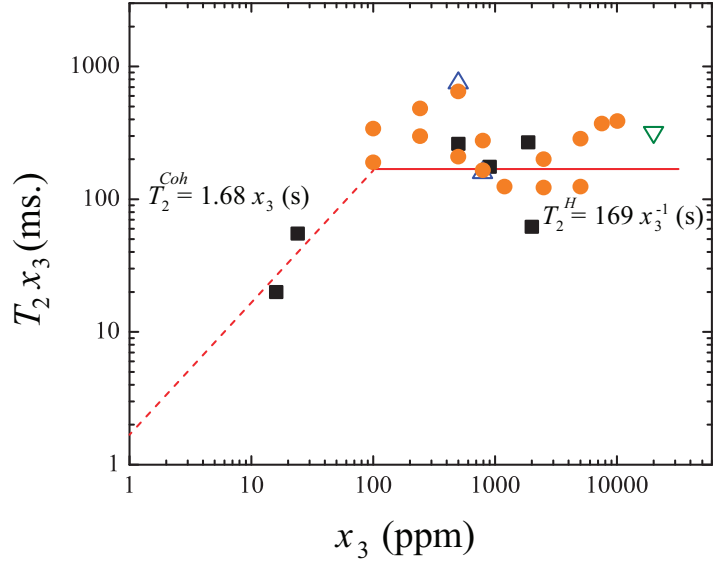}
\caption{ \label{fig: T2 function of x}
	(Color online) Observed $^3$He concentration dependence of the nuclear spin-spin relaxation time for dilute $^3$He in solid $^4$He for several concentrations reported by different authors.   The solid lines $T_2^L$ and $T_2^H$ are the predictions of the theories of Landesman\cite{Landesman1975137} and Huang {\it et al.}\cite{PhysRevB.11.3374}, respectively. $T_2^{Coh}$ is the value of $T_2$ for coherent diffusion where the mean separation is greater than the scattering length. See legend of Fig. 6 for definiton of symbols.   }
\end{figure}

First we consider the theoretical predictions for $T_2$ for small $x_3$. The values of the correlation time calculated by Landesman (Eqn. 10) that give a good fit to the $T_1$ data do not lead to values of $T_2$ that are in agreement with the experimental data. This discrepancy has been reviewed by Kim {\it et al.}\cite{KimLT26}  who showed that the detailed spectral density at low frequencies cannot be described by a single Lorentzian spectral density. A better approach to evaluating $T_2$ was given by Huang {\it et al.}\cite{PhysRevB.11.3374} who calculated the scattering of tunneling $^3$He impurities in the elastic crystal fields using detailed energy conservation  and found $T_2^H=1.69 \times 10^{-4}x_3$ s. for $x_3 >$  100 ppm. This result is compared with the experimental data and Landesman's theory in Fig.\~8. For very dilute concentrations, $x_3 <$ 100 ppm, the model of coherent diffusion as discussed above gives $T_2^{coh}=1.68$ s.

For these dilute samples the tunneling motion of the $^3$He impurities would lead to a temperature independent relaxation, but once again, we observed strong deviations from temperature independence below 175 mK. However, unlike the $T_1$ results which show a clean peak, there is considerable scatter in the $T_2$ data and the dependence observed is that of a small peak followed by a strong dip, and then at even lower temperature there is an order of magnitude drop at the phase separation temperature where  Fermi liquid droplets form. We now calculate what  would be expected for a simple superfluid phase transition. In terms of the spectral densities (at zero frequency) for the superfluid and normal components we have
\begin{equation}\frac{1}{T_{2Obs}} = \alpha(T)J_S(0) +[1-\alpha(T)]J_E(0)\end{equation}
while the fixed temperature independent tunneling term is
\begin{equation}\frac{1}{T_{2E}}=J_E(0)\end{equation}

\noindent We therefore find
\begin{equation} \frac{T_{2Obs}}{T_{2E}}= \frac{1}{b\alpha(T)+[1-\alpha(T)]} \end{equation} where $b=J_S(0)/J_E(0)=\tau_{S0}/{\tau_E} > 1$ and $\tau_E$ is the characteristic tunneling time in the normal solid. Because of the large value of $b$, the result for $T_{2Obs}$ does not look like the inverted image of $T_1$ (as one would naively expect) but is much wider. An example of an attempted fit for the adjustable parameters $b=4$ and $a_1=0.01$ is shown in comparison with experimental values of $T_2$ in Fig.\~9. Even allowing for the scatter of the data, this approach does not provide a good description of the temperature dependence of $T_2$. We should note that if the system is very inhomogeneous with different parts of the sample having different values of $b$ and $a_1$ one would expect a lumpy spectral density at low frequencies  and a  corresponding scatter in the results for $T_2$, but since $T_1$ measures the spectral weight at MHz frequencies, one could still observe a sharp peak for $T_1$ because it is the integrated weights of the low frequency components of the spectral densities that must be subtracted from the high frequency spectral density. In this sense the $T_1$ measurements are a much cleaner test for the existence of a phase transition that the $T_2$ measurements.

\begin{figure}[h]
\includegraphics*[width=0.55\textwidth]{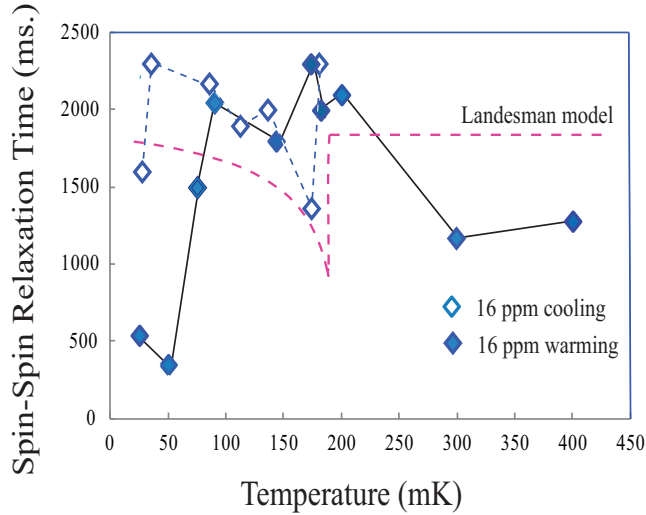}
\caption{ \label{fig: T2 Fit function of T}
	(Color online) Comparison of the observed temperature dependence of the nuclear spin-spin relaxation time $T_2$ for a sample with $x_3$ = 16  ppm with the dependence expected for a true phase transition at T=175 mK (broken purple line) as described in the text with the high temperature lmit given by Landesman's model. 
	 }
\end{figure}

\subsection{ Effect of lattice dynamics on the NMR relaxation rates.}

An alternative approach to interpreting the  NMR relaxation rates is to assume that as a result  of the tunneling of the $^3$He atoms there is an internal stress on the $^4$He lattice as the distortion around the $^3$He impurity moves with or attempts to move with the impurity as it tunnels from site to site. The mean frequency of this motion is $\omega_x=x^{4/3}K_0 \approx 3.5 \times 10^3$ Hz. We can surmise that the relaxation of the lattice due to this time dependent disturbance is long  and adds a bottleneck for energy exchange between the tunneling excitations and the thermal bath of phonons. The correlation function for the lattice operators $\tau(T)$ of Eq'n (3) can be written as
\begin{equation}
\langle\tau(t)\tau(0)\rangle=\langle\tau(0)^2\rangle_R exp(-i\omega_x+\tau_S^{-1})t
\end{equation}
where $\langle......\rangle_R$ represents an average over  lattice coordinates,  $\tau_S$ is the lattice averaged relaxation time which we assume corresponds to the Debye relaxation $\tau_S=\tau_0exp(E_0/T)$ used to interpret the results of the shear modulus measurements.\cite{PhysRevLett.104.195301,PhysRevB.79.214524,ElasticPropsBeamish09}  Integrating Eq'n (17, we find for the additional lattice relaxation time 
\begin{equation}\tau_x(T)=r_1\frac{\tau_0e^{E_0/T}}{1+(\omega_x\tau_0e^{E_0/T})^2}=\frac{r_1}{\omega_x}\frac{u}{1+u^2}\end{equation}

\noindent where $u=\omega_x\tau_0e^{(E_0/T)}$. $\tau_0$ and $E_0$ are adjustable parameters that we expect to be comparable to the 
values determined from the shear modulus experiments .{\cite{ElasticPropsBeamish09,Day_Beamish} The magnitude of the relaxation time $r_1$ can be estimated crudely using the golden rule with $r_1=\frac{4\pi}{\hbar}\sum_{E}|\langle E_1|K|E_2\rangle |^2\rho(E)\delta(E_1-E_2-E)$ where $K$ is the lattice distortion and the density of states $\rho(E)\sim E^2/E_D^2$. For $E_1-E_2=J$, the exchange energy, and $E_D$ the Debye energy we find $r_1/\omega_x \sim 4. 10^{-4}$.  In Fig.\~10 we show the fit to the relaxation times $T_1$, using $T_{1Obs}=T_{1E}+ \tau_x(T)$ for $\tau_0=8.3 \times 10^{-9}$ s. and $E_0=1.8$ K. These values are to be compared with those inferred from the shear modulus experiments,\cite{PhysRevLett.104.195301,PhysRevB.79.214524,ElasticPropsBeamish09} $\tau_0 = 2.3 \times 10^{-9}$ s. and $E_0=1.7$ K.\cite{PhysRevLett.104.195301} Although the values of $\tau_0$ and $E_0$ are a little  different from those deduced by Beamish {\it et al.},\cite{PhysRevB.79.214524,ElasticPropsBeamish09}  the fit is remarkably good given the approximations that have been made. This interpretation of the results is consistent with the observations of Sasaki {\it et al.}\cite{PhysRevB.81.214515,Sasaki2011} who followed the temperature dependence of the NMR signals of 10 ppm $^3$He in solid $^4$He for densities such that the phase separation forms solid clusters of $^3$He. While they were able to observe the solid clusters they were not able to detect the signal from isolated $^3$He atoms above the phase transition at low temperatures, and they attributed this to a long spin-lattice relaxation time  that lead to saturation   and signal loss at the temperatures of interest for exploring the anomalies of the $^4$He lattice.

\begin{figure}[h]
\includegraphics*[width=0.5\textwidth]{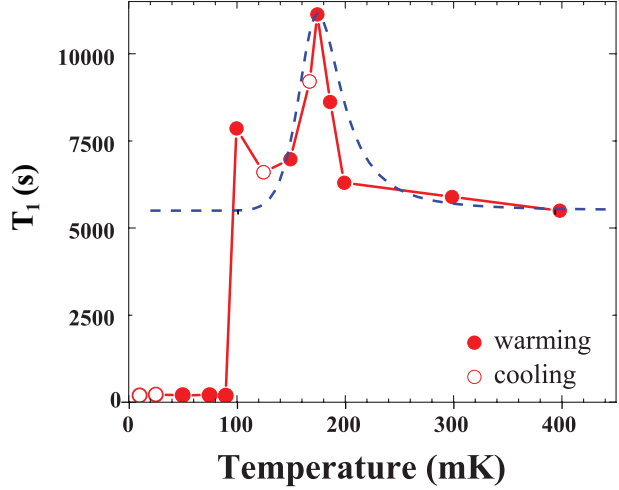}
\caption{ \label{fig: T1 Fit function of T}
	(Color online) Comparison of the observed temperature dependence of the nuclear spin-lattice relaxation time $T_1$ (for a sample with $x_3$ =  16 ppm) with the dependence calculated for  lattice relaxation due to the motion of the distortion of the lattice around impurities, using parameters inferred from the measurements of the shear modulus. }
	
\end{figure}

\subsection{Effect of dislocations on relaxation times}

 The most significant effect of the presence of dislocations on the NMR studies is that the $^3$He atoms will pin the dislocation lines leaving some of the $^3$He atoms located on the dislocation lines and the lines themselves essentially immobile at low temperatures. The number of $^3$He atoms that are expected to be affected by this is very small. The localization of the $^3$He atoms would lead to three distinct components of the NMR signals and their relaxation: (i) the contribution from atoms fixed on the dislocation lines, (ii) the interactions between fixed localized $^3$He atoms with those tunneling freely in the solid, and (iii) the interactions between pairs where both are free to move in the lattice. The latter has been the focus of the discussions above. For those atoms that are essentially fixed the line shapes would be very broad leading to short values of $T_2$ and very long values for $T_1$. The echoes would have more than one contribution and if the numbers of fixed atoms were significant the echo would have a central peak, corresponding to the spin-spin interactions between immobile atoms. We have not observed such a peak or a deviation from a single exponential decay for the dilute samples down to the lowest temperatures studied.
 
This result is not surprising when one considers the number of dislocations present even in crystals of average quality. Paalanen {\it et al.}\cite{Paalanen1981} and Iwasa {\it et al.}\cite{Iwasaetal1978,Iwasa2010} show that the dislocation density $\Lambda$ and the dislocation network length $L_N$ are related by $\Lambda L_N^2=0.2$ and $\Lambda \cong 10 ^6$ cm$^{-2}$ with $L_N \sim (3-6)\times 10^{-4}$ cm. $\sim 10^4 a_0$. The concentration of $^3$ He atoms that would saturate the dislocation network is therefore  estimated as $x_{3s}\sim (\frac{a_0}{L_N})^2 \sim 10^{-8}$. The dislocation density would need to be much larger than that expected for even average quality crystals\cite{Iwasaetal1978,Iwasa2010} if the pinned $^3$He atoms were to be observable.

In addition to dislocation lines, Balibar has pointed out \cite{BalibarPrivate} that grain
boundaries are also likely to bind $^3$He impurities - possibly affecting NMR
relaxation times in a temperature-dependent manner.  A key parameter
determining the importance of grain-boundary binding is the typical grain
size $L_G$. We have no direct information on $L_G$ for our blocked-capillary
samples, although, for example, $L_G \approx 0.5$~mm was observed by Franck
{\it et al.} \cite{Franck} in annealed helium films.

The grain size $L_G$ controls any possible effect of grain-boundary binding
on the NMR response in two distinct ways: (a) the maximum concentration of
$^3$He that can be bound at grain boundaries is of order $a_0/L_G$, assuming
full monolayer coverage with bound $^3$He, and (b) the time scale for $^3$He
atoms to freely diffuse on and off grain boundaries is of order $L_G^2/D$,
where $D$ is the diffusion coefficient.  For $L_G = 0.5$~mm we calculate
$a_0/L_G =0.7$ ppm, much less than the $^3$He concentrations used in the present
experiments. For a diffusion constant $D=3. 10^{-6}$ cm$^2$/s.\cite{1982JLTP...47..289A} we find $L_G^2/D \approx $10$^3$ s., much shorter than the relaxation
times $T_1$ that we observe. The latter implies that the collisions between the $^3$He impurities in the bulk sample dominates the relaxation process. Therefore, it appears that the grain size in
our samples would need to be much smaller than was observed in films in Ref.[59] for the binding of $^3$He atoms at grain boundaries to be the cause of the
$T_1$ anomalies we observe.


\subsection{NMR Relaxation for Droplets}

For all samples studied (except the 24 ppm sample) a sharp phase transition is observed at low temperatures below which the NMR amplitudes are independent of temperature (see Fig. 4) as expected for the well-known phase separation into Fermi liquid droplets.\cite{ PhysRevB.39.4083} The observed phase separation temperatures are in good agreement with the values calculated by Edwards and Balibar.\cite{PhysRevB.39.4083} The rate of formation of the droplets is very slow (typically 2-10 hours)\cite{HuanDroplets} and great care must be taken to ensure that equilibrium is reached before measuring amplitudes and relaxation times near the phase separation.
After phase separation, the relaxation times are observed to become temperature independent.\cite{HuanDroplets,Kingsley1998} The observed relaxation times are consistent with a relaxation that occurs at the wall of degenerate Fermi liquid droplets. Huan {\it et al.}\cite{HuanDroplets} showed that the relaxation time was given by $T_1=(N_{core}/N_{wall})\tau_X$ where $N_{core}$ and $N_{wall}$ are the number of atoms in the core of the droplet and in the wall respectively. $\tau_X$ is the intrinsic relaxation time at the wall due to the $^3$He tunneling is given by $\tau_X=J_3/M_2$ where $J_3$ is the tunneling rate at the wall and $M_2$ is the NMR second moment. This estimate leads to values of $T_1 \sim$ 170 s. in good qualitative agreement with the observations (see Fig. 5).

\section{Conclusion}

Measurements of the nuclear spin-lattice relaxation times for very dilute $^3$He concentrations have shown the existence of pronounced peaks in the relaxation times at the same temperatures as those for which anomalies are observed in torsional oscillator and shear modulus measurements of solid $^4$He. Less well-defined variations are observed for the nuclear spin-spin relaxation times. The detailed temperature dependences do not fit a model in which there is a well-defined phase transition to a  supersolid or superfluid state where the critical fluctuations can induce dramatic changes in the relaxations times with critical exponents for the temperature dependence near the transition temperature. The observations are best described by the introduction of an additional relaxation process in series with the usual tunneling-relaxation process and caused  by the response of the lattice to the motion of the lattice distortions around the tunneling impurity atoms. The characteristic parameters for this model, the tunneling rate and the lattice excitation energy are remarkably close to those values deduced from measurements of the shear modulus.
Further studies are needed at lower $^3$He concentrations  and lower magnetic fields to to create a better separation between the phase separation temperature and the  temperatures for which the NMR and other anomalies are observed.

\begin{acknowledgments}
	This research was carried out at the NHMFL High B/T Facility which is supported by NSF Grant DMR 0654118 and by the State of Florida.
	This project was supported in part by an award from the Collaborative Users Grant Program of the NHMFL.
	We gratefully acknowledge useful discussions with  Sebastian Balibar, Moses Chan, Brian Cowan, Alan Dorsey, Izumi Iwasa and Pradeep Kumar.
\end{acknowledgments}

\bibliography{3Hein4He-SSK-PRB}
\end{document}